\documentclass[MSNbibl,nameyear,dvips]{arxstspdf}
\usepackage{flushend}
\usepackage{stfloats}

\volume{26}
\issue{2}
\pubyear{2011}
\firstpage{206}
\lastpage{209}
\doi{10.1214/11-STS338B}
\referstodoi{10.1214/10-STS338}

\makeatletter
\renewcommand{\epsilon}{\varepsilon}
\newcommand{\E}{\mathrm{E}}
\makeatother

\begin{document}
\begin{frontmatter}

\title{Discussion of ``Objective Priors: An~Introduction for Frequentists'' by~M.~Ghosh}
\runtitle{Discussion}
\pdftitle{Discussion of Objective Priors: An Introduction for Frequentists by M. Ghosh}

\begin{aug}
\author{\fnms{Trevor} \snm{Sweeting}\corref{}\ead[label=e1]{trevor@stats.ucl.ac.uk}}
\runauthor{T. Sweeting}

\affiliation{University College London}

\address{Trevor Sweeting is Professor of
Statistical Science, Department of Statistical Science,
University College London, London WC1E 6BT, United Kingdom  \printead{e1}.}

\end{aug}



\end{frontmatter}

The paper by Ghosh provides a useful introduction to the main ideas
underlying objective priors and how these ideas might profitably be
used by frequentist statisticians, both at a theoretical and practical
level.  The aspects likely to be of most interest to this group of
statisticians are those concerning probability matching, allowing valid
frequentist procedures to be derived via a formal Bayesian analysis.
But they should also be interested in priors that arise from
decision-theoretic considerations, not least since the consideration of
risk criteria, such as mean squared error for estimation or operating
characteristic function for testing, is ubiquitous in the frequentist
approach.  As pointed out by the author, at a theoretical level the
shrinkage argument, which I~have also used extensively in the past,
provides a~neat way of deriving frequentist asymptotic results.

My discussion will focus on an examination of the main criteria that
have been used to obtain objective priors and, partly related to this,
the extent to which the theory and practical application can be
extended to more complex scenarios.  Before launching into this I would
just like to comment on the commonly used term ``objective'' in the
present context.  As soon becomes apparent in this field, there is an
array of possible criteria available for the development of objective
priors, some of which depend on a specific choice of parameterization,
and there may be no unique solution even for a given criterion.  Thus
the choice quickly ceases to be purely objective.  My own preference is
to use the term\vadjust{\eject} ``nonsubjective,'' which indicates that the prior is
detached from subjective beliefs about parameters but which does not
impart such a strong sense of broad agreement as to what the prior
should be in any particular case.

\section{Comparison of Criteria}

First, a general point about alternative criteria for the development
of objective priors.  I have a strong preference for criteria that
would lead to the use of properly calibrated subjective priors whenever
they are available, so that the consideration of objective priors in
some sense generalizes a property of a fully subjective Bayesian
approach.  In a sense this is true of probability matching since this
leads to (approximately) correct coverage of posterior regions in
hypothetical repeated sampling.  This in turn implies that these
regions will also be calibrated over repeated {\it use}, as would
automatically be the case if a~properly elicited subjective prior were
to be used.   The same cannot be said for moment matching in the sense
described in Section 5.2; there seems nothing in this criterion that
would lead one to use a~subjective prior when available.

Similarly, consideration of a proper scoring rule in a~decision-theoretic approach would indicate the use of an elicited
subjective prior whenever one is available.  As a consequence, I would
be uneasy using a decision-theoretic criterion that was not based on a
proper scoring rule. For example, it does seem surprising that, even in
the scalar parameter case, Jeffreys' prior turns out not to be optimal
under the distance measure (3.13) with $\beta=-1$.  The problem is
that, unlike the Bernardo criterion that arises when $\beta=0$ (see
later), none of these distance measures corresponds to an average
regret based on some primitive loss function that produces a (negative)
score when data $x$ are observed and a prior predictive distribution
$\pi(x)$ is adopted.  So there seems to be no obvious sense in which we
would recover a~subjective prior distribution whenever one is
available.

Although there is some reference to predictive pro\-bability matching in
Sections 5 and 6, the paper is largely a review of objective priors
obtained via parametric criteria, which usually require a focus on one
or more specified parameters of interest.  This has certainly been the
most popular area of study and, as a~technical device for obtaining
frequentist procedures, it performs a useful function.  However, the
focus on parameters is a cause for concern for many Bayesian
statisticians.  Such approaches normally require a specific choice of
parameters of interest, such as in quantile probability matching or the
construction of group reference priors.  The idea that an analysis
should be redone when the spotlight turns to alternative sets of
parameters is disturbing.  In particular, in complex real-world
applications there will potentially be many parametric functions of
interest.  An alternative to quantile matching is higher-order matching
for highest posterior density or other regions, which may not require a~specific choice of interest parameters.  However, there is an infinite
variety of ways in which a region can be chosen.  Indeed, in the scalar
parameter case, given \textit{any} prior it is possible to choose the
region in such a way that higher-order matching is achieved (Severini,
\citeyear{Sev93}; Sweeting, \citeyear{Swe99}).

An alternative approach is to study the behavior of predictive
distributions.  This is appealing as the parameterization then becomes
irrelevant.  Just as in the parametric case one can consider predictive
probability matching (Datta, Ghosh and Mukerjee, \citeyear{DatGhoMuk00}; Severini, Mukerjee and Ghosh,
\citeyear{SevMukGho02}) and predictive risk (Komaki, \citeyear{Kom96}; Sweeting, Datta  and Ghosh, \citeyear{SweDatGho06}), and
Ghosh has contributed to both of these areas.  In the former case the
criterion~(4.23) is replaced by the following.  Let $Y$ be a~future
observation from the model and let $y(\pi,\alpha)$ denote the
$(1-\alpha)$-quantile of the predictive distribution of~$Y$ based on
the prior $\pi$.  If it is also the case that
\[
\operatorname{pr}\{Y>y(\pi,\alpha)|\theta\}=\alpha+O(n^{-r}),\vspace*{2pt}
\]
then we have predictive probability matching; typically~$r$ will be 2
here. In the latter case we can consider the regret when the prior
$\pi$ is adopted and $\theta$ is the true parameter value.  Adopting
the logarithmic scoring rule $-\log\pi(y|x)$,  which is the unique
local proper scoring rule, this has the general form\vspace*{3pt}
%
\begin{equation}\label{yx}
d_{Y|X}(\theta,\pi)=\E^\theta\biggl[\log\biggl\{\frac{f(Y|X,\theta)}{\pi(Y|X)}\biggr\}\biggr].
\end{equation}
Priors that attempt to control this risk might be considered to be more
`general purpose' than priors that require the specification of certain
parametric functions.

Having used a sensible broad criterion to obtain a~prior, one could
then go on to investigate its parametric properties.  For example,
there may be more than one prior that produces the same (low)
predictive risk and the choice between these priors might be made on
the basis of a particular interest parameterization.  In Examples 1 and
2 of the paper the right Haar prior $\pi(\mu,\sigma)\propto\sigma^{-1}$
is exactly predictive probability matching and also arises as a minimax
prior under (\ref{yx}) (Liang and Barron, \citeyear{LiaBar04}).   We can then see
that, for example, it is exactly probability matching when the interest
parameter is $\mu$ or $\sigma$ and second-order probability matching
when $\theta=\mu/\sigma$ is the interest parameter, as shown in Example
2 (continued).

It is instructive to compare the above predictive risk criterion with
the basic reference prior approach of Bernardo (\citeyear{Ber79}, \citeyear{Ber05}).  The
reference prior criterion in Section 3.1 is maximization of the
Kullback--Leibler divergence between the prior and posterior
distributions.  As shown by Clarke and Barron (\citeyear{ClaBar94}), this is
equivalent to finding the  minimax solution under the regret
%
\begin{equation}\label{x}
d_X(\theta,\pi)=\E^\theta\biggl[\log\biggl\{\frac{f(X|\theta)}{\pi(X)}\biggr\}\biggr].
\end{equation}
Note that (\ref{x}) is based on the proper scoring rule $-\log\pi(x)$.
This may be contrasted with (\ref{yx}), which is based on the proper
scoring rule $-\log\pi(y|x)$, as suggested by Geisser in his discussion
of Bernardo (\citeyear{Ber79}).  The former is based on scoring the prior
predictive distribution, which is arguably less relevant than the
posterior predictive distribution on which the latter is based.  We are
not so much interested in predicting the data already observed as new
data yet to be observed.  This distinction is reminiscent of model
fitting, where it is the fit to as yet unobserved data that is more
relevant than the fit to observed data.  Note also that working in
terms of the posterior predictive distribution avoids problems of
impropriety of the prior, requiring only that $\pi(x)<\infty$.  Thus,
to continue the discussion of Example 1 in the paper, in contrast to
the predictive criterion (\ref{yx}), Jeffreys' prior emerges as  the
minimax solution under~(\ref{x}), whereas it is inadmissible under~(\ref{yx}).

In more complex examples (\ref{yx}) involves a complicated function
that includes components of skewness and curvature of the model.
However, it is argued in Sweeting, Datta and Ghosh (\citeyear{SweDatGho06}) that it
is more appropriate to consider the regret
%
\begin{equation}\label{yxtau}
d_{Y|X}(\tau,\pi)=\E\biggl[\log\biggl\{\frac{\tau(Y|X)}{\pi(Y|X)}\biggr\}\biggr],
\end{equation}
where the expectation is taken over the joint distribution of $X$ and
$Y$ under the prior $\tau$.  This is because we are not so much
interested in comparing the performance of $\pi$ with that in a
lower-dimensional submodel at a fixed parameter value as comparing its
performance with that of other nondegenerate prior distributions for
the current model.  Moreover, when an elicited prior $\tau$ is
available criterion (\ref{yxtau}) will lead us to use this prior.  An
asymptotic analysis of (\ref{yxtau}) and the adoption of a minimax
criterion, for example, produces sensible priors in specific examples.
Another appealing aspect is that the asymptotic predictive criterion
does not depend on the amount of prediction.

\vspace*{-1pt}
\section{More Complex Models}
\vspace*{-1pt}

Some of the most important and challenging applications of the day,
such as environmental science, biomedicine, neuroscience and genomics,
demand lar\-ge, sophisticated and often high-dimensional models. The
results in Section 4 of the paper on first- and second-order matching
priors are mathematically attractive, but there is clearly a need to
explore the extent to which these results can be profitably used in
more complex models.  As the author points out in Section~6, objective
priors have been successfully developed for a number of more complex
problems.  However, there remains a need for semi-automated procedures
so that suitable ``safe'' default priors can be developed rapidly for
arbitrary model structures.   Major difficulties include the difficulty
or impossibility of obtaining a closed form expression for Fisher's
information and, even if this is possible, of solving the required
partial differential equations.  Levine and Casella (\citeyear{LevCas03}) proposed an
algorithm for the implementation of probability matching priors for a
single interest parameter in the presence of a~single nuisance
parameter.  However, the implementation requires a substantial amount
of computing time.  An alternative approach is outlined in Sweeting
(\citeyear{Swe05}), where it is shown that suitable data-dependent priors can be
developed in some cases.  Staicu and Reid (\citeyear{StaRei08}) proposed an elegant
analytic solution based on higher-order approximation of the marginal
posterior distribution.  It seems to me, however, that some form of
data-driven approach will be the only viable way to extend probability
matching ideas to general frameworks.

Apart from computational difficulties, the major theoretical difficulty
of all the approaches to objective prior construction that rely on
sample size asymptotics is the potential breakdown of the theory in
high-dimensional parameter spaces.  In some cases it may be possible to
identify directions in the parameter space about which the data are
relatively uninformative.  This can be conveniently explored, for
example, via an eigenanalysis of the observed information matrix.
Although the model is high-dimensional, most of the variation of the
likelihood may take place on a lower-dimensional manifold of the
parameter space.  This means, of course, that the model is close to
being non-identifiable, which causes difficulties if the parameters
themselves are of direct interest.  However, this may be amenable to
analysis using a predictive approach.  If a parameter only enters
weakly in the model, then the predictive distribution should not depend
critically on the prior chosen for that parameter and asymptotic theory
should apply in such cases.

Although versions of probability matching priors and reference priors
in nonregular cases have been investigated by Ghosal (\citeyear{Gho97}, \citeyear{Gho99}) and
Berger, Ber\-nardo and Sun (\citeyear{BerBerSun09}), it will be a major challenge to develop
multidimensional priors in an automatic way when some aspects of the
model are regular and others nonregular.

I suspect that the application of objective priors for high-dimensional
problems will be of greater interest to Bayesian than to frequentist
statisticians.  Given the difficulties of deriving such priors in these
cases, the frequentist may well abandon this route and explore
alternative simulation-based approaches.  On the other hand, a suitable
high-dimensional prior is essential for the Bayesian statistician to
operate at all.  Yet the greater the dimension of the model the less
likely it is that reliable prior information will be available on all
the parameters, let alone on their mutual dependencies.  Furthermore,
as noted earlier, it is less likely that there will be just one or two
parameters of interest, so I believe that the quest will focus more on
the identification of safe, general purpose priors that allow the
inclusion of subjective information when available, rather than on
priors tailored to specific parameters.  If this ambition is realized,
then the resulting priors should be thought of as no more than
``reference'' priors, in the broad sense of the word, and should not
replace the need for sensitivity analysis.

\section{Some Other Difficulties}

Many Bayesian statisticians remain sceptical about the need for
objective priors to represent ignorance and a common practice is to
utilize proper but diffuse priors instead.  However, care has to be
taken that the tail behavior of such priors is not too thin, otherwise
the prior may have the unexpected effect of dominating the likelihood.
Consider a random sample from $N(\mu,\sigma^2)$.  Suppose that $\mu$
and $\sigma^2$ are taken to be a priori independent with normal and
inverse Gamma distributions, respectively.  How diffuse should these
distributions be and how sensitive are the results to these choices?
Specifically, suppose that $X_i\sim N(\mu, \phi^{-1})$, where $\phi$ is
the precision parameter, and $\mu,\phi$ are a priori independent with
$\mu\sim N(0,c^{-1}), \phi\sim\operatorname{Gamma}(a,b)$.  Suppose we observe
data $529.0, 530.0, 532.0, 533.1, 533.4, 533.6,\break 533.7, 534.1, 534.8,
535.3$.  Take $a=b=c=\epsilon$.  What is the effect of the choice of
$\epsilon$?  The value $c=0.001$ is not small enough: the
``noninformative prior'' dominates the likelihood and the mean of the
marginal posterior of $\mu$ is close to zero.  Effectively, this
happens because the normal tail of the prior for~$\mu$ is thinner than
the Student $t$-tail of the integrated likelihood of $\mu$.  The value
$c=0.0002$ is also not sufficiently small, although if a Gibbs sampler
starting near the sample values is run, then it will not detect the
problem at all until after a large number of ite\-rations and it will
appear from trace plots as if the sampler has converged.  A value of
$c$ less than 0.0001 is needed for the likelihood to dominate the
prior.  If we run into such problems in simple models like this, then
there has to be a great deal of concern for higher-dimensional models.
So objective priors do matter; it is virtually impossible to reliably
elicit a high-dimensional prior distribution and there are pitfalls
associated with using vague but proper priors.

Yet another difficulty arises when the likelihood does not tend to zero
at the boundary of the parameter space.  In that case an improper prior
may lead to an improper posterior, forcing the use of a~proper prior.
The objective selection of such a prior is likely to be problematic.
An example is the dispersion parameter in a Dirichlet process mixture
model.  Some authors simply set the hyperparameters in a Gamma prior to
be very small, but clearly this requires great care as we know that in
the limit we will obtain an improper posterior.

\section{Concluding Remarks}

I do think that frequentist interest in Bayesian statistics should be
rather more than simply its potential use as a device to obtain valid
frequentist procedures.  When there is some concern about the priors
adopted, Bayesians will often ``look over their shoulder'' at
frequentist properties, if only to check that the prior is not
producing some anomalous behavior (cf. Example 3 in the paper).
Likewise, frequentist statisticians should find it useful to do the
same, possibly to provide an indication that they are not falling
seriously foul of the conditionality principle, or possibly to see to
what extent their confidence statements have direct probability
interpretations.  Finally, I would like to thank the author for his
interesting review of this area and for stimulating me to think a
little more about the basis for the construction of objective priors
and the challenges that confront this field of research.



\begin{thebibliography}{15}

\bibitem[\protect\citeauthoryear{Berger, Bernardo and Sun}{2009}]{BerBerSun09}
\begin{barticle}[mr]
\bauthor{\bsnm{Berger},~\bfnm{James~O.}\binits{J.~O.}},
  \bauthor{\bsnm{Bernardo},~\bfnm{Jos{\'e}~M.}\binits{J.~M.}} \AND
  \bauthor{\bsnm{Sun},~\bfnm{Dongchu}\binits{D.}}
(\byear{2009}).
\btitle{The formal definition of reference priors}.
\bjournal{Ann. Statist.}
\bvolume{37}
\bpages{905--938}.
\bid{doi={10.1214/07-AOS587}, issn={0090-5364}, mr={2502655}}
\end{barticle}
\endbibitem

\bibitem[\protect\citeauthoryear{Bernardo}{1979}]{Ber79}
\begin{barticle}[mr]
\bauthor{\bsnm{Bernardo},~\bfnm{Jose-M.}\binits{J.-M.}}
(\byear{1979}).
\btitle{Reference posterior distributions for {B}ayesian inference (with discussion)}.
\bjournal{J. Roy. Statist. Soc. Ser. B}
\bvolume{41}
\bpages{113--147}.
\bid{issn={0035-9246}, mr={0547240}}
\end{barticle}
\endbibitem

\bibitem[\protect\citeauthoryear{Bernardo}{2005}]{Ber05}
\begin{bincollection}[mr]
\bauthor{\bsnm{Bernardo},~\bfnm{Jos{\'e}~M.}\binits{J.~M.}}
(\byear{2005}).
\btitle{Reference analysis}.
In \bbooktitle{Bayesian Thinking: Modeling and Computation}.
\bseries{Handbook of Statist.}
\bvolume{25}
(\beditor{\bfnm{D.~K.}\binits{D.~K.}~\bsnm{Dey}} \AND
\beditor{\bfnm{C.~R.}\binits{C.~R.}~\bsnm{Rao}, eds.})
\bpages{17--90}.
\bpublisher{North-Holland, Amsterdam}.
\bid{mr={2490522}}
\end{bincollection}
\endbibitem

\bibitem[\protect\citeauthoryear{Clarke and Barron}{1994}]{ClaBar94}
\begin{barticle}[mr]
\bauthor{\bsnm{Clarke},~\bfnm{Bertrand~S.}\binits{B.~S.}} \AND
  \bauthor{\bsnm{Barron},~\bfnm{Andrew~R.}\binits{A.~R.}}
(\byear{1994}).
\btitle{Jeffreys' prior is asymptotically least favorable under entropy risk}.
\bjournal{J.~Statist. Plann. Inference}
\bvolume{41}
\bpages{37--60}.
\bid{doi={10.1016/0378-3758(94)90153-8}, issn={0378-3758}, mr={1292146}}
\end{barticle}
\endbibitem

\bibitem[\protect\citeauthoryear{Datta, Ghosh and Mukerjee}{2000}]{DatGhoMuk00}
\begin{barticle}[mr]
\bauthor{\bsnm{Datta},~\bfnm{Gauri~Sankar}\binits{G.~S.}},
  \bauthor{\bsnm{Ghosh},~\bfnm{Malay}\binits{M.}} \AND
  \bauthor{\bsnm{Mukerjee},~\bfnm{Rahul}\binits{R.}}
(\byear{2000}).
\btitle{Some new results on probability matching priors}.
\bjournal{Calcutta Statist. Assoc. Bull.}
\bvolume{50}
\bpages{179--192}.
\bid{issn={0008-0683}, mr={1843620}}
\end{barticle}
\endbibitem

\bibitem[\protect\citeauthoryear{Ghosal}{1997}]{Gho97}
\begin{barticle}[mr]
\bauthor{\bsnm{Ghosal},~\bfnm{S.}\binits{S.}}
(\byear{1997}).
\btitle{Reference priors in multiparameter nonregular cases}.
\bjournal{Test}
\bvolume{6}
\bpages{159--186}.
\bid{doi={10.1007/BF02564432}, issn={1133-0686}, mr={1466439}}
\end{barticle}
\endbibitem

\bibitem[\protect\citeauthoryear{Ghosal}{1999}]{Gho99}
\begin{barticle}[mr]
\bauthor{\bsnm{Ghosal},~\bfnm{Subhashis}\binits{S.}}
(\byear{1999}).
\btitle{Probability matching priors for non-regular cases}.
\bjournal{Biometrika}
\bvolume{86}
\bpages{956--964}.
\bid{doi={10.1093/biomet/86.4.956}, issn={0006-3444}, mr={1741992}}
\end{barticle}
\endbibitem

\bibitem[\protect\citeauthoryear{Komaki}{1996}]{Kom96}
\begin{barticle}[mr]
\bauthor{\bsnm{Komaki},~\bfnm{Fumiyasu}\binits{F.}}
(\byear{1996}).
\btitle{On asymptotic properties of predictive distributions}.
\bjournal{Biometrika}
\bvolume{83}
\bpages{299--313}.
\bid{doi={10.1093/biomet/83.2.299}, issn={0006-3444}, mr={1439785}}
\end{barticle}
\endbibitem


\bibitem[\protect\citeauthoryear{Levine and Casella}{2003}]{LevCas03}
\begin{barticle}[mr]
\bauthor{\bsnm{Levine},~\bfnm{Richard~A.}\binits{R.~A.}} \AND
  \bauthor{\bsnm{Casella},~\bfnm{George}\binits{G.}}
(\byear{2003}).
\btitle{Implementing matching priors for frequentist inference}.
\bjournal{Biometrika}
\bvolume{90}
\bpages{127--137}.
\bid{doi={10.1093/biomet/90.1.127}, issn={0006-3444}, mr={1966555}}
\end{barticle}
\endbibitem


\bibitem[\protect\citeauthoryear{Liang and Barron}{2004}]{LiaBar04}
\begin{barticle}[mr]
\bauthor{\bsnm{Liang},~\bfnm{Feng}\binits{F.}} \AND
  \bauthor{\bsnm{Barron},~\bfnm{Andrew}\binits{A.}}
(\byear{2004}).
\btitle{Exact minimax strategies for predictive density estimation, data
  compression, and model selection}.
\bjournal{IEEE Trans. Inform. Theory}
\bvolume{50}
\bpages{2708--2726}.
\bid{doi={10.1109/TIT.2004.836922}, issn={0018-9448}, mr={2096988}}
\end{barticle}
\endbibitem

\bibitem[\protect\citeauthoryear{Severini}{1993}]{Sev93}
\begin{barticle}[mr]
\bauthor{\bsnm{Severini},~\bfnm{Thomas~A.}\binits{T.~A.}}
(\byear{1993}).
\btitle{Bayesian interval estimates which are also confidence intervals}.
\bjournal{J. Roy. Statist. Soc. Ser. B}
\bvolume{55}
\bpages{533--540}.
\bid{issn={0035-9246}, mr={1224415}}
\end{barticle}
\endbibitem

\bibitem[\protect\citeauthoryear{Severini, Mukerjee and
  Ghosh}{2002}]{SevMukGho02}
\begin{barticle}[mr]
\bauthor{\bsnm{Severini},~\bfnm{Thomas~A.}\binits{T.~A.}},
  \bauthor{\bsnm{Mukerjee},~\bfnm{Rahul}\binits{R.}} \AND
  \bauthor{\bsnm{Ghosh},~\bfnm{Malay}\binits{M.}}
(\byear{2002}).
\btitle{On an exact probability matching property of right-invariant priors}.
\bjournal{Biometrika}
\bvolume{89}
\bpages{952--957}.
\bid{doi={10.1093/biomet/89.4.952}, issn={0006-3444}, mr={1946524}}
\end{barticle}
\endbibitem

\bibitem[\protect\citeauthoryear{Staicu and Reid}{2008}]{StaRei08}
\begin{barticle}[mr]
\bauthor{\bsnm{Staicu},~\bfnm{Ana-Maria}\binits{A.-M.}} \AND
  \bauthor{\bsnm{Reid},~\bfnm{Nancy~M.}\binits{N.~M.}}
(\byear{2008}).
\btitle{On probability matching priors}.
\bjournal{Canad. J. Statist.}
\bvolume{36}
\bpages{613--622}.
\bid{doi={10.1002/cjs.5550360408}, issn={0319-5724}, mr={2532255}}
\end{barticle}
\endbibitem

\bibitem[\protect\citeauthoryear{Sweeting}{1999}]{Swe99}
\begin{barticle}[mr]
\bauthor{\bsnm{Sweeting},~\bfnm{Trevor~J.}\binits{T.~J.}}
(\byear{1999}).
\btitle{On the construction of {B}ayes-confidence regions}.
\bjournal{J. R. Stat. Soc. Ser. B Stat. Methodol.}
\bvolume{61}
\bpages{849--861}.
\bid{doi={10.1111/1467-9868.00206}, issn={1369-7412}, mr={1722243}}
\end{barticle}
\endbibitem

\bibitem[\protect\citeauthoryear{Sweeting}{2005}]{Swe05}
\begin{barticle}[mr]
\bauthor{\bsnm{Sweeting},~\bfnm{Trevor~J.}\binits{T.~J.}}
(\byear{2005}).
\btitle{On the implementation of local probability matching priors for interest
  parameters}.
\bjournal{Biometrika}
\bvolume{92}
\bpages{47--57}.
\bid{doi={10.1093/biomet/92.1.47}, issn={0006-3444}, mr={2158609}}
\end{barticle}
\endbibitem

\bibitem[\protect\citeauthoryear{Sweeting, Datta and Ghosh}{2006}]{SweDatGho06}
\begin{barticle}[mr]
\bauthor{\bsnm{Sweeting},~\bfnm{Trevor~J.}\binits{T.~J.}},
  \bauthor{\bsnm{Datta},~\bfnm{Gauri~S.}\binits{G.~S.}} \AND
  \bauthor{\bsnm{Ghosh},~\bfnm{Malay}\binits{M.}}
(\byear{2006}).
\btitle{Nonsubjective priors via predictive relative entropy regret}.
\bjournal{Ann. Statist.}
\bvolume{34}
\bpages{441--468}.
\bid{doi={10.1214/009053605000000804}, issn={0090-5364}, mr={2275249}}
\end{barticle}
\endbibitem

\end{thebibliography}
\end{document}